\documentclass[apjl]{emulateapj}
\usepackage{psfig,amsfonts,amsmath,graphicx,natbib,apjfonts}
\usepackage{ulem}

\def\new#1 {{\bf #1 }}
\def\cut#1 {\sout{#1} }
\def\cfa{1}
\def\ra#1#2#3{#1$^{\rm h}$#2$^{\rm m}$#3$^{\rm s}$}
\def\dec#1#2#3{$#1^\circ#2'#3''$}

\shorttitle{VLBA Observations of Ultracool Dwarfs}
\shortauthors{Forbrich \& Berger}

\begin{document}

\title{The First VLBI Detection of an Ultracool Dwarf: Implications
for the Detectability of Sub-stellar Companions}

\author{
J.~Forbrich\altaffilmark{\cfa}
\& E.~Berger\altaffilmark{\cfa}
}

\altaffiltext{\cfa}{Harvard-Smithsonian Center for Astrophysics, 60
Garden Street, Cambridge, MA 02138}

\begin{abstract} We present milliarcsecond-resolution radio very long
baseline interferometry (VLBI) observations of the ultracool dwarfs
TVLM\,513--46546 (M8.5) and 2MASS J00361617+1821104 (L3.5) in an
attempt to detect sub-stellar companions via direct imaging or reflex
motion.  Both objects are known radio emitters with strong evidence
for periodic emission on timescales of about 2 and 3 hours,
respectively.  Using the inner seven VLBA antennas, we detect unresolved
emission from TVLM\,513--46546 on a scale of 2.5 mas ($\sim 50$
stellar radii), leading to a direct limit on the radio emission
brightness temperature of $T_B\gtrsim 4\times 10^5$ K.  However, with
the higher spatial resolution afforded by the full VLBA
we find that the source appears to be marginally and
asymmetrically resolved at a low S/N ratio, possibly indicating that 
TVLM\,513--46546 is a
binary with a projected separation of $\sim 1$ mas ($\sim 20$ stellar
radii).  Using the
7-hr baseline of our observation we find no astrometric shift in
the position of TVLM\,513--46546, with a $3\sigma$ limit of about 0.6
mas. This is about 3 times larger than expected for an equal mass
companion with a few-hour orbital period. Future monitoring of its
position on a range of timescales will provide the required
astrometric sensitivity to detect a planetary companion with a mass of
$\sim 10$ M$_J$ in a $\gtrsim 15$ d ($\gtrsim 0.06$ AU) orbit, or with
a mass of $\sim 2$ M$_J$ in an orbit of $\gtrsim 0.5$ yr ($\gtrsim
0.3$ AU).
\end{abstract}


\section{Introduction}

In recent years unexpectedly strong radio emission has been detected
from very low mass stars and brown dwarfs (hereafter, ultracool
dwarfs), revealing that these objects are capable of generating and
dissipating kG-strength magnetic fields (e.g.,
\citealt{ber01,ber06,hal08}).  The radio luminosity remains nearly
uniform from early-M to mid-L dwarfs, even though other magnetic
activity indicators (H$\alpha$ and X-rays) decline by about two orders
of magnitude over the same spectral type range (e.g.,
\citealt{whw+04,ber06}).  Thus, radio observations provide a
particularly powerful probe of the magnetic field properties.  

The radio emission from several ultracool dwarfs also exhibits a
periodicity that matches the stellar rotation velocity ($v\,{\rm sin}
i$), thereby pointing to a large-scale, axisymmetric, and stable
magnetic field configuration on timescales of hours to years
\citep{ber05,hal06,hal07,ber09}.  These conclusions are also supported
by observations of Zeeman broadening in FeH molecular lines
\citep{rei07} and time-resolved optical spectropolarimetry
\citep{dmp08,mdp08}.

Since ultracool dwarfs are fully convective, the solar-type
$\alpha\Omega$ dynamo, which operates at the shearing interface
between the radiative and convective zones \citep{par55}, cannot be
responsible for generating and amplifying the inferred fields.
However, recent numerical simulations suggest that large scale
axisymmetric fields can still be generated in fully convective
objects, at least for conditions that roughly correspond to mid-M
dwarfs ($M\sim 0.3$ M$_\odot$; \citealt{bro08}).  Simulations that
correspond to the conditions in ultracool dwarfs are challenging and
have not been investigated so far.  As a result, observational
constraints on the scale, geometry, and origin of the fields are
essential.  The spectroscopic Zeeman techniques are currently of
limited utility for the dim and generally rapidly-rotating ultracool
dwarfs, since they require very high signal-to-noise ratios and have
reduced sensitivity at high rotation velocities due to line broadening
\citep{rei07}.  Detailed radio observations are thus crucial.

Here we present the first very long baseline interferometry (VLBI)
observations of ultracool dwarfs, aimed at providing further
constraints on their magnetic field properties.  In particular, these
observations have sufficient angular resolution to detect astrometric
shifts of a few tenths of a milliarcsecond that may be exerted by a
sub-stellar companion with a period of a few hours.  They also allow us
to directly image a radio-emitting companion down to milliarcsecond
scales.  The presence of such putative companions may enhance the
magnetic activity through direct or tidal interaction.  We detect the
well-studied M8.5 dwarf TVLM\,513-46546 with the VLBA, the first such
detection for any ultracool dwarf.  The detectability of this object
with the VLBA suggests that long-term astrometric monitoring could
reveal the presence of a planetary-mass companion on a $\gtrsim 15$ d
orbit.  Such a detection would also usher in a new era of extrasolar
planet studies through radio astrometry.

\section{Targets and Observations}
\label{sec:targets}

The M8.5 dwarf TVLM\,513--46546 (hereafter TVLM513--46) is located at
a distance of 10.6 pc \citep{dah02}.  It was first detected at radio
wavelengths by \citet{ber02} who found persistent emission and a
circularly polarized flare lasting about 15 min at 8.5 GHz.
Subsequently, \citet{ost06} detected quiescent radio emission at 4.9
and 8.5 GHz, but did not detect any flares.  \citet{hal06} found
rotational modulation in the radio emission with a periodicity of
$\sim 2$ hr, and subsequently, \citet{hal07} detected periodic bursts
of circularly polarized radio emission with a period of 1.96 hr.  Most
recently, \citet{ber08} confirmed the presence of polarized flares
combined with quiescent emission in simultaneous multi-wavelength
observations (radio, X-ray, and H$\alpha$).  These authors also found
weak X-ray emission and periodic H$\alpha$ emission.  Thus, the
current picture of the radio emission from TVLM513--46 is one
involving different activity states; quiescent radio emission appears
to always be present ($200-400$ $\mu$Jy at 8.5 GHz), but variability
and periodic bursts have not been detected in all observations.

We also observed 2MASS J00361617+1821104 (hereafter, 2M0036+18), an
L3.5 dwarf located at a distance of 8.8 pc \citep{dah02}.  This object
has also been previously detected at radio wavelengths: \citet{ber05}
found strongly variable and periodic radio emission with a periodicity
of 3 hr, which was later confirmed by \citet{hal08}.  The average 8.5
GHz flux density measured previously was about 135 $\mu$Jy
\citep{ber05}.

\subsection{Radio Observations}

TVLM513--46 and 2M0036+18 were observed simultaneously with the Very
Long Baseline Array (VLBA), the Very Large Array (VLA), and the Green
Bank Telescope (GBT) on 2008 March 30.  Four hours of on-source time
in the VLBI observations were obtained at 05:45--12:32 UT for
TVLM513--46 and 13:40--20:35 UT for 2M0036+18; the VLA observations
covered the entire time range.  All data were analyzed with the
Astronomical Image Processing System (AIPS).

\subsubsection{VLA}
\label{sec:vla}

VLA observations were obtained at a frequency of 8.5 GHz with 26
antennas in the C configuration.  We used the standard continuum setup
with $2\times 50$ MHz bands and full polarization.  Flux calibration
was obtained using J1331+305, while J1455+2131 (for TVLM513--46) and
J0019+2021 (for 2M0036+18) were used as complex gain calibrator
sources.  The bootstrapped flux densities for these two sources were
0.081 Jy and 0.866 Jy, respectively.  About once every 50 min
additional scans of secondary calibrators were obtained, resulting in
corresponding gaps in the target source light curves.

\subsubsection{VLBA+VLA+GBT}

The VLBI observations included the 10-antenna VLBA, the GBT, and the
VLA as a phased array.  Phase-referenced observations were carried out
with a data rate of 256 Mbit s$^{-1}$ in dual polarization, using
2-bit sampling.  Eight base-band channels of 8 MHz bandwidth each
covered an aggregate bandwidth of 32 MHz.  The correlator dump time
was 2 s. The complex gain calibrators J1455+2131 and J0019+2021 were
located at distances of $1.8^\circ$ and $4.4^\circ$, respectively.

Unfortunately, the phase calibrators were not sufficiently bright to
phase up the VLA between scans on the target sources, and the phased
VLA data had to be removed from the VLBI experiment.  Similarly, the
GBT experienced bad weather, and the telescope had to be stowed for
nearly 2.5 hr due to freezing rain during the second half of the
experiment.  As a result of the poor phases we also removed the GBT
from the VLBI experiment.  Finally, of the ten VLBA antennas, we
removed Hancock (Hn), which experienced heavy snow fall and hence
elevated system temperatures.  We therefore separately analyze the VLA
and VLBA data sets.

\section{Results}

The VLA data allow us to determine the flux densities of our target
sources, to construct light curves, and to constrain the search areas
for counterparts on VLBI scales.

\subsection{VLA observations}

\begin{figure}
\centering
\includegraphics*[angle=-90,width=\linewidth,bb=54 55 325 620]{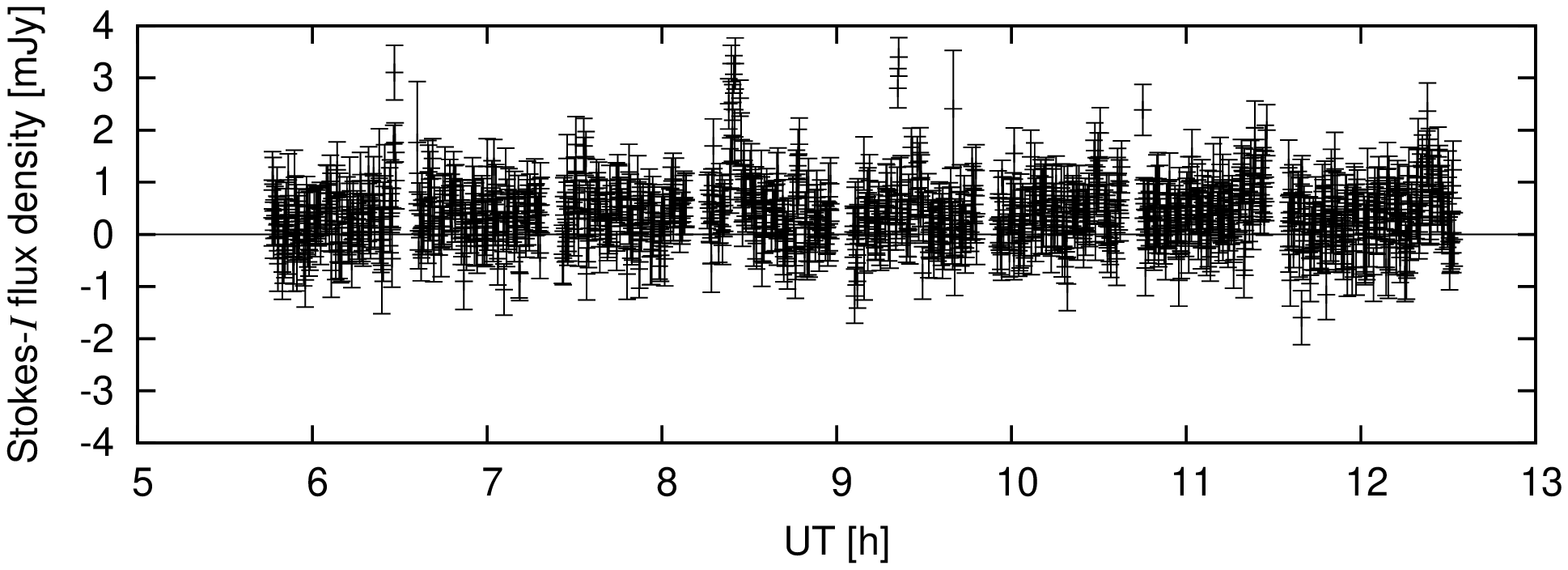}
\includegraphics*[angle=-90,width=\linewidth,bb=150 55 367 620]{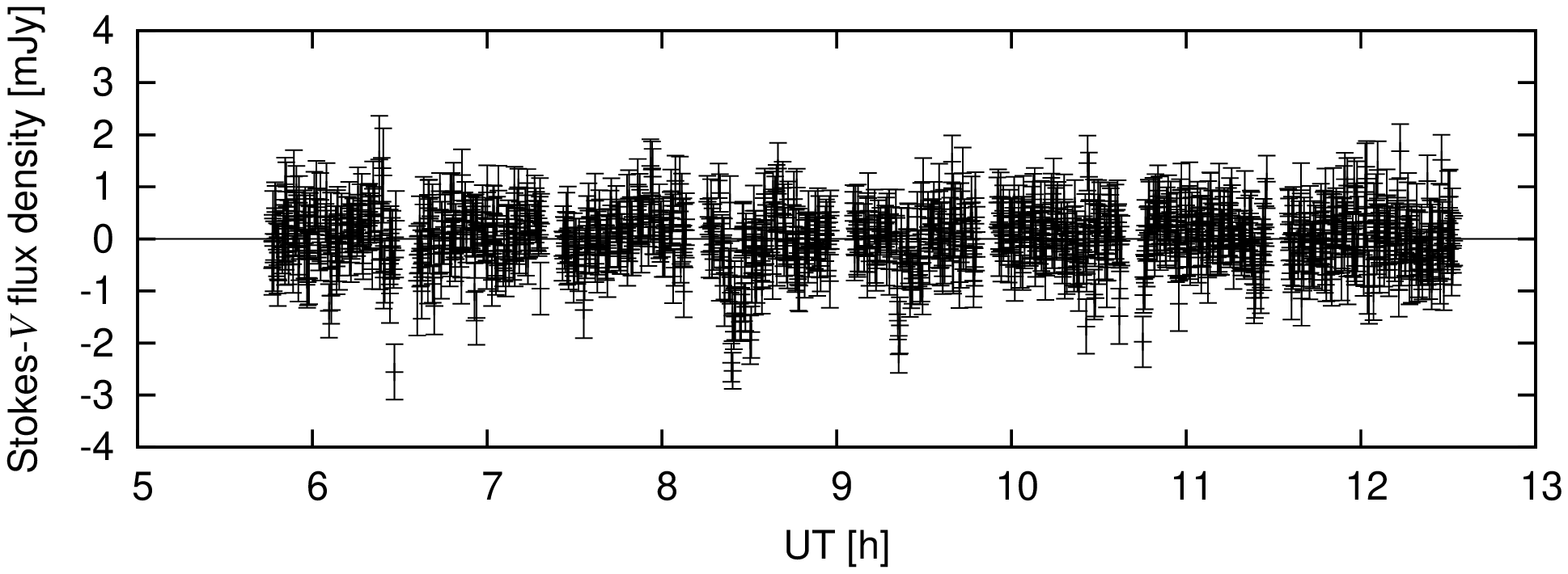}
\caption{VLA 8.5 GHz light curve of TVLM513--46 with a time resolution
of 10 s.  The upper panel shows the total intensity (Stokes $I$), and
the lower panel shows the circularly polarized flux (Stokes $V$);
negative values indicate left-handed circular polarization.  For
better visibility, all points that have error bars more than three
times larger than the average error were discarded.
\label{dwf1501_lc2}}
\end{figure}

Both sources are detected in the VLA observations.  TVLM513--46 has a
flux density of $539\pm 19$ $\mu$Jy, and is located at a position of
RA=\ra{15}{01}{08.162} ($\pm 0.002$ s), Dec=\dec{+22}{50}{01.673}
($\pm 0.035''$).  The synthesized beam size is $2.3''\times 2.2''$.
No circular polarization is detected in the integrated data, with a
$3\sigma$ limit of $r_c\lesssim 9\%$.  The light curve reveals that
the emission is composed of a quiescent component and short-duration
flares lasting for only a few minutes but with circular polarization
approaching $\sim 100\%$ (Figure~\ref{dwf1501_lc2}).  The most
prominent flares are detected at 08:25 and 09:22 UT.  During the first
two hours, when no flares are detected, the flux density is $445\pm
28$ $\mu$Jy.

\begin{figure}
\centering
\includegraphics[width=6.0in]{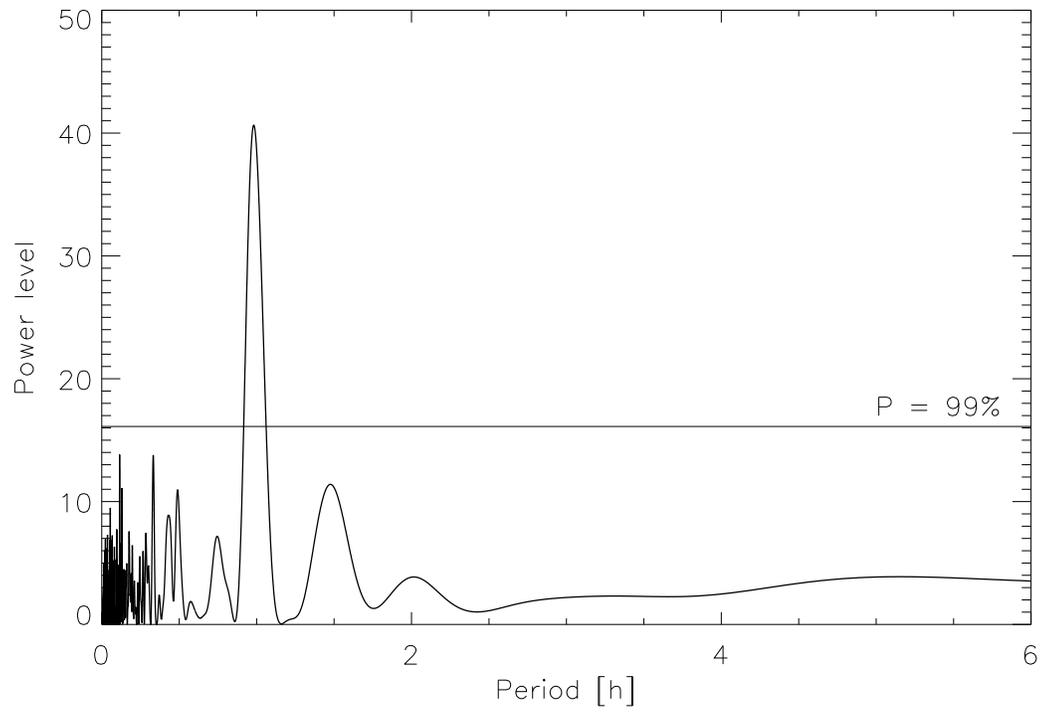}
\caption{Lomb-Scargle periodogram for the VLA 8.5 GHz light curve of
TVLM513--46.  The horizontal line denotes the power level
corresponding to a false alarm probability of 0.01.  We find a
significant peak at a period of 58.8 min.
\label{scarglefig}}
\end{figure}

In light of previous results indicating that TVLM513--46 exhibits
periodic radio variability, we conducted a periodicity search using
the Lomb-Scargle periodogram \citep{sca82}; Figure~\ref{scarglefig}.
We find a signal at a period of 58.8 min, exactly half of the
previously reported period.  Surprisingly, no obvious signal is
detected at the harmonic of 117.6 min, even though three cycles at
that periodicity fit into the observation.  The 58.8-min period is the
only peak in the periodogram above the critical level of about 16 that
corresponds to a false alarm probability of 0.01. Note that the false
alarm probability does not take into account the flux density errors
of the light curve.

We also detect 2M0036+18 with a flux density of $144\pm 22$ $\mu$Jy,
similar to previous detections (\S\ref{sec:targets}).  No significant
variability was detected in the light curve.

\subsection{VLBA observations}
\label{sec:vlba}

\begin{figure}
\centering
\includegraphics*[angle=-90,width=6.0in, bb=70 110 540 670]{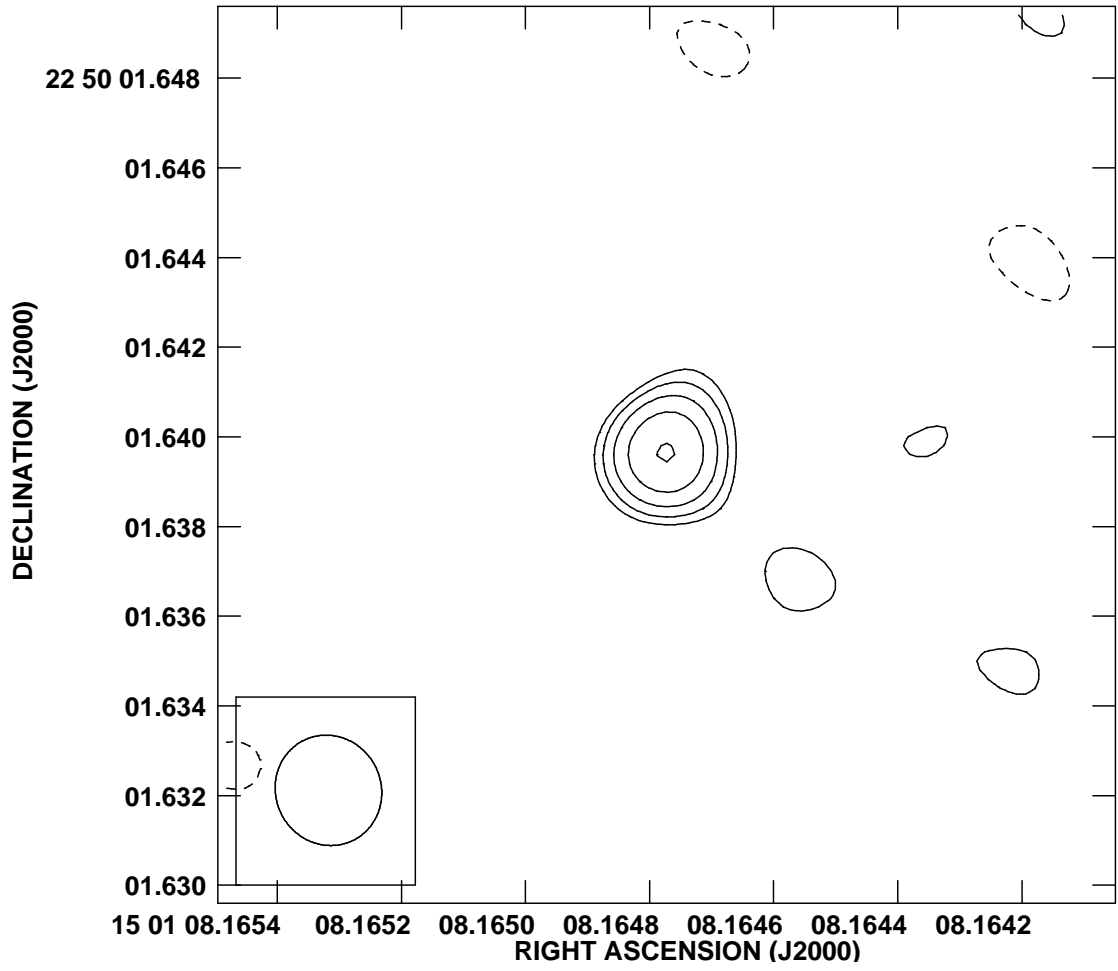}
\caption{VLBA detection of TVLM513--46 using the inner seven antennas.
Contour lines are --2,2,$2\sqrt 2$,4,$4\sqrt 2$,8 $\sigma$, where
$\sigma=48$ $\mu$Jy; the synthesized beam size is indicated in the
lower left corner.
\label{dwf1501_vlbi_1}}
\end{figure}

TVLM513--46 is detected with the VLBA, the first such detection of any
ultracool dwarf.  To locate the source in a data set using the inner
seven VLBA antennas we shifted the phase center of the $uv$ data to
the nominal VLA position and created a map covering $0.2''\times
0.2''$, encompassing the VLA position uncertainty ($1\sigma=45$ mas).
We find a single robust source at the $8\sigma$ significance level,
which persists with changes in the imaging parameters and source
detection routines (AIPS SAD and JMFIT).  The source is located 46 mas
(i.e., $1\sigma$) from the nominal VLA position.  We therefore
identify this source as the VLBA counterpart of TVLM513--46 and shift
the phase center to this position for subsequent analysis.  The source
is located at a position of RA=\ra{15}{01}{08.1647696} ($\pm
0.0000077$ s), Dec=\dec{+22}{50}{01.63968} ($\pm 0.00011''$), where
the errors are statistical only (Figure~\ref{dwf1501_vlbi_1}).

With a synthesized beam size of $2.5\times 2.3$ mas (${\rm PA}=
26^\circ$) the source appears to be unresolved; it has a peak
brightness of $420\pm 48$ $\mu$Jy beam$^{-1}$ and an integrated flux
density of $360\pm 75$ $\mu$mJy (the VLA and VLBA fluxes overlap at
the $2\sigma$ level).  At the distance of TVLM513--46, the beam size
corresponds to a spatial resolution of about 0.026 AU, or about 50
times larger than the expected radius of TVLM513--46.

\begin{figure}
\centering
\includegraphics*[angle=-90,width=6.0in, bb=70 110 540 670]{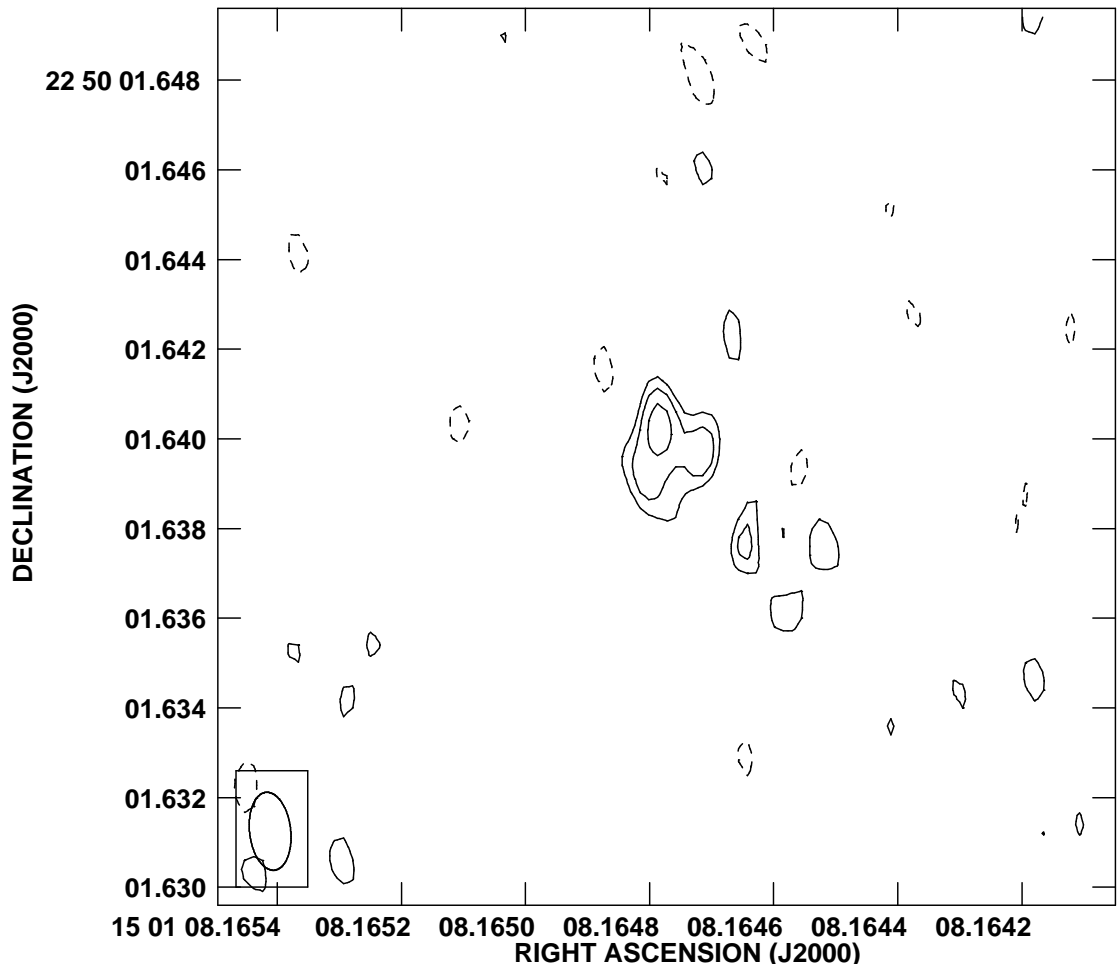}
\caption{VLBA detection of TVLM513--46 using the entire VLBA (minus
Hn).  Contour lines are --2,2,$2\sqrt 2$,4,$4\sqrt 2$ $\sigma$, where
$\sigma=50$ $\mu$Jy; the synthesized beam size is indicated in the
lower left corner.  The source appears to be marginally resolved and
asymmetric at this resolution, possibly indicating that TVLM513--46 is
a roughly equal-mass binary system.
\label{dwf1501_vlbi_2}}
\end{figure}

To follow up on this detection with an increased angular resolution,
we added the outer VLBA antennas located on Mauna Kea and St.~Croix.
The resulting synthesized beam size is $1.7\times 0.9$ mas (${\rm PA}
=6^\circ$).  At this resolution, the S/N ratio decreases, and the peak brightness of TVLM513--46 is
$230\pm 47$ $\mu$Jy beam$^{-1}$, and its integrated flux density of
$540\pm 150$ $\mu$Jy has a larger error bar since the source no longer
appears to be well-described by a single unresolved Gaussian
component.  Instead, it appears asymmetric and resolved
(Figure~\ref{dwf1501_vlbi_2}).  This indicates a marginally resolved
source on the longest baselines, but the significance is too low to
warrant a firm conclusion.  The synthesized beam corresponds to a
source size of about 20 stellar radii, larger than expected for the
magnetic field scale (e.g., \citealt{lin83}).  We therefore conclude
that if TVLM513--46 is indeed resolved, then this likely indicates a
binary system with two radio-emitting sources and a projected
separation of about 1 mas.  In this case, since radio emission has
only been detected to date from objects with spectral type earlier
than L3.5 \citep{ber06}, the binary would likely be close to equal
mass.

We note that 2M0036+18 was not detected in the VLBA data due to
insufficient sensitivity.

\section{Implications of the VLBA Detection}

Since TVLM513--46 is unresolved on a scale of about 2.5 mas, we can
place a firm lower limit on the brightness temperature of its
quiescent emission of $\gtrsim 4\times 10^5$\,K.  This is not
sufficiently high to {\it directly} rule out thermal emission (e.g.,
\citealp{dul85}).  To constrain the brightness temperature to values
above $10^7$ K, indicative of non-thermal emission, the angular
resolution will have to be $\lesssim 0.4$ mas.  We note that other
considerations (expected size of the magnetic field, radio spectral
index) point to non-thermal emission (e.g., \citealp{ost06}).

The sub-milliarcsecond astrometry of TVLM513--46 that is available
thanks to its detectability with the VLBA underscores the potential
for a VLBI astrometric companion search.  Since TVLM513--46 is already
at the stellar/sub-stellar transition, such a companion would be a
sub-stellar object (brown dwarf or planet).  \citet{bow09} have
recently used such VLBA astrometry to search for companions to nearby
early-M dwarfs.  They detected five M dwarfs with spectral types M1-M5
in one to three epochs with the VLBA down to a $5\sigma$ sensitivity
limit of about 0.5 mJy.  Their astrometric residuals compared to
optical measurements of the parallax and proper motion were about 0.2
mas, similar to the astrometric accuracy we achieved here for
TVLM513--46.

The signal-to-noise ratio of our detection only allows us to search
for an astrometric shift during the 7-hour observation when separately
imaging the first and second halves of the data set.  Using data from
the inner 7 VLBA antennas we find a positional difference of $0.47\pm
0.37$ mas, consistent with no significant shift.  We note that the
expected maximum astrometric shift (equal-mass companion) for an orbit
of a few hours is about $0.2$ mas, indicating that VLBA observations
alone (with $1\sigma\approx 0.2$ mas) cannot detect the presence of
such a companion at the brightness level of TVLM 513--64.  However,
more sensitive combined VLBA+GBT observations may provide the required
signal-to-noise ratio to probe the presence of an equal-mass companion
in a few-hour orbit (corresponding to a separation of $\sim {\rm few}$
stellar radii).  Moreover, if TVLM513--46 is indeed a roughly equal
mass binary with a projected separation of about 1 mas ($\gtrsim 1$ d
orbit) we expect an easily detectable astrometric signature of $\sim 1$ 
mas on a 1 d timescale.

\begin{figure}[ht!]
\centering
\includegraphics[width=4.0in,angle=-90]{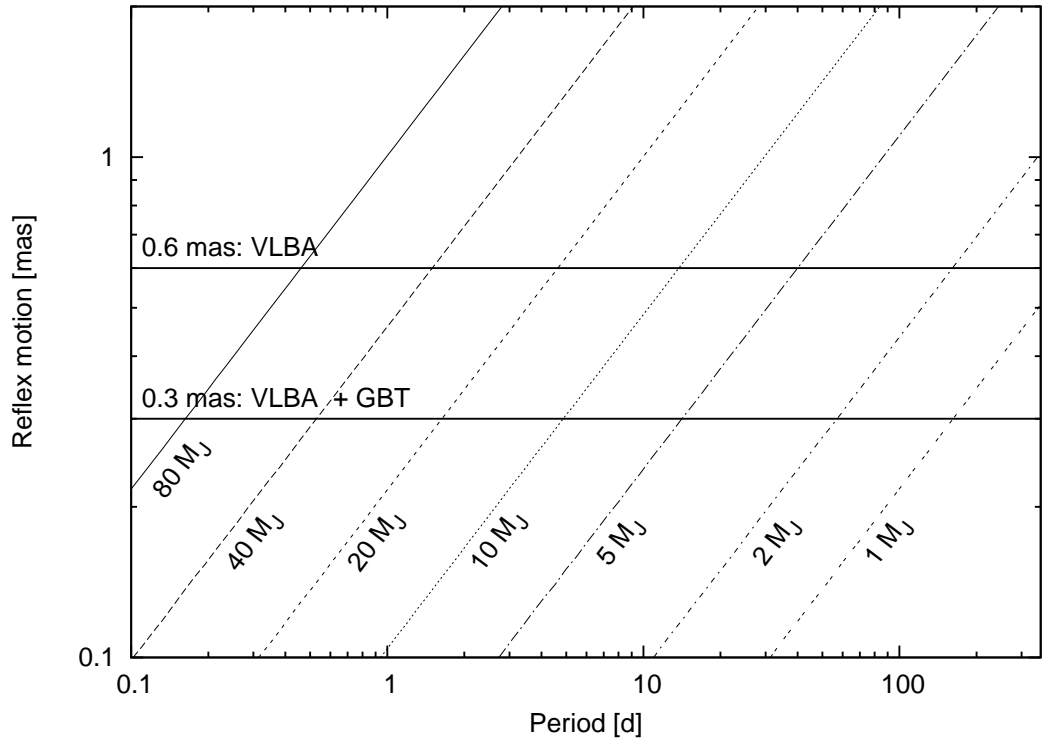}
\caption{Predicted maximum reflex motion for companions with masses of 80, 40, 20, 10, 5, 2, and 1 M$_J$ (left to right), as a function of the
corresponding orbital period, assuming that TVLM513--46 is a single
star.  While a single observation with the VLBA+GBT has the potential
to directly detect a radio-emitting (and hence roughly equal mass)
companion down to a scale of $\sim 1$ mas (corresponding to a $\sim 1$
d orbit), observations spaced over a year can detect a sub-stellar
companion down to a few Jupiter masses via reflex motion.  If
TVLM513--46 is indeed an equal-mass resolved binary on a
milliarcsecond scale, then we should be able to further constrain the
system with the detection of astrometric reflex motion, expected to be
at a level of about 1 mas.
\label{exoplanetfig}}
\end{figure}

Equally important, a planetary-mass companion to TVLM513--46 can be
detected on longer timescales, since a wider orbit leads to a larger
astrometric signal.  In Figure~\ref{exoplanetfig} we plot the maximum
reflex motion induced by a companion orbiting TVLM513--46 with a range
of masses ($1-40$ M$_J$) and orbital periods ($0.1$ d to 1 yr).  We
find that at $3\sigma\approx 0.6$ mas for VLBA observations similar to
the one presented here, we can detect a 10 M$_J$ companion with an
orbital period of $\gtrsim 15$ d ($\gtrsim 0.06$ AU), or a 2 M$_J$
companion with an orbital period of $\gtrsim 0.5$ yr ($\gtrsim 0.3$
AU).  Using the combined VLBA+GBT, with an expected astrometric
uncertainty of $3\sigma\approx 0.3$ mas (due to a higher
signal-to-noise ratio and a higher fraction of longer baselines), we
will be able to detect a 1 M$_J$ planetary companion in a $\gtrsim
0.5$ yr orbit.  We note that these limits are appropriate for the case
where TVLM513--46 is a single star.  If the object is an equal mass
binary, as may be indicated by the results of direct imaging, the
search for planetary-mass companions will be more challenging.

\section{Conclusions and Future Prospects}

Using high angular resolution radio VLBI observations, we clearly
detect the ultracool dwarf TVLM513--46 on an unprecedented small scale
of about 2.5 mas or about 50 stellar radii.  The source is unresolved
on this scale, but it may be marginally resolved on a $\sim 1$ mas
scale when using the longest VLBA baselines, although the S/N ratio
is low.  Given the corresponding
physical scale of about 20 stellar radii, this may point to a binary
system with two radio-emitting sources in a $\gtrsim 1$ d orbit.

Beyond the possibility of a spatially resolved binary, a search for an
astrometric shift in the position of the source due to the
gravitational influence of a close companion yields a $3\sigma$ limit
of $\lesssim 0.6$ mas, about 3 times larger than the expected shift
for an equal-mass companion in a few-hour orbit.  Since we lost the
phased-VLA and GBT data, we expect that future VLBI observations will
provide the requisite astrometric accuracy to search for an equal-mass
companion in such a tight orbit.  These future observations will also
allow us to determine whether the marginally-resolved emission from
TVLM513--46 is indeed due to an equal mass companion with a projected
separation of about 20 stellar radii.  If this is indeed the case, the
expected maximum reflex motion is about 1 mas, easily detectable
with VLBA+GBT observations that span several days.

Equally important, the fact that TVLM513--46 is detected with the VLBA
opens the possibility to astrometrically search for a planetary-mass
companion with a $\gtrsim 10$ d orbit.  Monitoring of TVLM513--46 with
the VLBA or VLBA+GBT on timescales of days to months to years will
allow us to probe the existence of companions with masses of $\sim
1-10$ M$_J$ and with orbits of $\sim 15$ d to $\sim 1$ yr.  These
observations may lead to the first detection of an extrasolar planet
via radio astrometry.

Thus, the use of VLBI observations opens up a wide companion parameter
space for TVLM513--46, with detection via direct imaging for a roughly
equal-mass radio-emitting companion down to a scale of $\sim 20$
stellar radii, and detection via reflex motion on scales as small as a
few stellar radii for an equal mass companion, and scales of $\gtrsim
0.1$ AU for a planetary mass companion.  This first VLBA detection
also paves the way for similar observations of known and future
radio-emitting ultracool dwarfs in search of companions.

\acknowledgments{We thank Michael Rupen (NRAO), Mark Reid (CfA) and 
Lorant Sjouwerman (NRAO) for assistance and helpful advice concerning the 
VLBI data. The National Radio Astronomy Observatory is a facility of the 
National Science Foundation operated under cooperative agreement by 
Associated Universities, Inc.}


\clearpage

\clearpage

\clearpage

\clearpage

\clearpage

\end{document}